\documentclass[prl,twocolumn]{revtex4}

\usepackage{amssymb,amsmath}
\newcommand{\ket}[1]{|#1 \rangle}
\newcommand{\bra}[1]{\langle #1 |}
\newcommand{\braket}[2]{\langle #1| #2 \rangle}

\begin{document}

\title{Encoding secret information in measurement settings}

\author{Syed~M~Assad}
\affiliation{Centre for Quantum Computation and Communication Technology, Department of Quantum Science, The Australian National University, Canberra, Australia}

\author{Amir~Kalev}
\affiliation{Center for Quantum Information and Control, University of New Mexico, Albuquerque, New Mexico 87131-0001, USA}

\begin{abstract}
Secure communication protocols are often formulated in a paradigm where the message is encoded in measurement outcomes. In this work we propose a rather unexplored framework in which  the message is encoded in measurement settings rather than in their outcomes. In particular,  we study two different variants of such secure communication protocols in which the message alphabet corresponds  to measurement settings of mutually unbiased bases. 
\end{abstract}

\maketitle

\section{Introduction}\label{intro}

Cryptographic protocols are designed to exchange secret messages
between two parties, usually referred to as Alice and Bob. Bennett and
Brassard \cite{bb84} first demonstrated that the fundamental
laws of quantum mechanics can play a central role in the security of a properly
formulated protocol. Their seminal work has pushed forward the field of quantum information theory and triggered a vast study of quantum cryptography.  Numerous variations and generalizations of their scheme have been suggested,  e.g.,~\cite{ekert91,bb92,bruss98,englert04}.

In a recent work, Kalev, Mann and Revzen~\cite{Kalev2013}  showed how the disturbance induced by a nonselective measurement can be used to establish a communication channel. They presented a protocol in which two parties  can establish a communication channel using the choice of measurement settings (rather than their outcomes) as signal. It was understood that the proposed protocol is not secure~\cite{Kalev2013,Xie2013} and left as an open question whether the protocol can be used for secret sharing. In this work, we address that question and demonstrate two extensions of this protocol that proved to be secure.
These extensions deviate from the common framework of existing quantum cryptographic protocols based on the original Bennett and Brassard protocol where the secret information is encoded in states of a quantum system through a measurement outcome. To our knowledge, this work presents the first quantum cryptography proposal in which the secret information is encoded in measurement settings instead of measurement outcomes.

From the onset, we admit that our proposal is experimentally challenging to implement and seems to offer no advantage over the original Bennett and Brassard protocol. We believe that the Bennett and Brassard protocol is superior to this and most later discrete variable protocols due to its simplicity. The aim of this modest work is not to overthrow existing quantum cryptographic protocols and their rich history. Instead, we merely wish to explore new possibilities for secure quantum communication, which may in turn lead to new insights on quantum cryptography.

\section{Communication by choice of measurement}
Before presenting the secure variants, we recall the original communication protocol that uses the choice of measurement basis as a signal \cite{Kalev2013}. Alice and Bob first define a set of  $d+1$ mutually 
unbiased bases (MUBs)  \cite{ivanovic81,tal,durt10} in a prime dimension $d$. The first basis, denoted by $b=\ddot{0}$, is the computational 
basis $\{\ket{n}\}_{n=0}^{d-1}$. The remaining $d$ bases, parametrized by $b=0,1,\ldots,d-1$, are given in terms 
of the computational basis by \cite{tal}
\begin{equation}
\label{eqn:mub}
 \ket{m;b} = \frac{1}{\sqrt{d}} \sum_{n=0}^{d-1} \ket{n} \omega^{b n^2 -2 n m}\;,
\end{equation}
for $m,b=0,1,\ldots,d-1$ where
\begin{align}
\omega&=i \textrm{ for } d=2 \textrm{ and }\\
\omega&=e^{\frac{2\pi i}{d}} \textrm{ for } d>2\;.
\end{align}
No further classical communication takes place. 

Next Alice prepares one of the $d^3$ two-qudit maximally entangled state \cite{rev1,rev2}:
\begin{equation}
\label{eqn:crs}
\ket{c,r;s}_{1,2} = \frac{1}{\sqrt{d}} \sum_{n=0}^{d-1} \ket{n}_1 \ket{c-n}_2 \omega^{s n^2 -2 r n}\;,
\end{equation}
with $c,r,s=0,1,\ldots, d-1$. Throughout this letter, the arithmetics is modulo $d$. For each $s$, these states form an orthonormal basis for the Hilbert space of the two qudits. Alice keeps qudit 1 and sends qudit 2 to Bob. After this, Bob chooses $b$, the signal that he wishes to communicate, 
and measures qudit 2 in the MUB $\{\ket{m;b}_{m=0}^{d-1}\}$. The outcome of the measurement is irrelevant and therefore Bob may choose to disregard it. Finally, Bob returns the 
measured qudit 2 to Alice who then measures both qudits 1 and 2 in the basis of preparation $\{\ket{c',r';s}_
{1,2}\}_{c',r'=0}^{d-1}$. From the outcome $c',r'$, Alice can infer the value of $b$ from the following table:
\begin{equation}
\begin{array}{l l l l}
\label{dec_tab}
&c' \neq c &\rightarrow& b=s-\frac{r-r'}{c-c'}\;,\\
r' \neq r,&c' = c &\rightarrow& b= \ddot{0}\;,\\
r' = r, &c' = c &\rightarrow& \textrm{inconclusive}\;.
\end{array}
\end{equation}
The probability of an inconclusive outcome is $\frac{1}{d}$. As an example, the case for $d=2$ is summarized in probability table~\ref{tab:d2}.

\begin{table}
\centering
\begin{tabular}{cccccc}
\toprule
\multicolumn{2}{c}{Bob's choice}&\multicolumn{4}{c}{Alice's outcomes $\ket{c',r'}$}\\
%\cline{1-2}
$b$&basis&$\ket{0,0}$&$\ket{0,1}$&$\ket{1,0}$&$\ket{1,1}$\\
\colrule
$\ddot{0}$&$Z$&$\frac{1}{2}$&$0$&$\frac{1}{2}$&$0$\\
0&$X$&$\frac{1}{2}$&$0$&$0$&$\frac{1}{2}$\\
1&$Y$&$\frac{1}{2}$&$\frac{1}{2}$&$0$&$0$\\
\botrule
\end{tabular}
\caption{Communication by choice of measurement basis protocol for $d=2$ when Alice prepares the two qubit state with $c=r=s=0$. This state corresponds to one of the four bell states. The three mutually unbiased bases Bob can choose to measure in correspond to the $X$, $Y$ and $Z$ axes of the Bloch sphere. When Alice performs her measurement in the Bell basis, an outcome $\ket{0,0}$ gives her an inconclusive result. The other three outcomes inform her of Bob's measurement choice $b$.}
\label{tab:d2}
\end{table}

This protocol is vulnerable to an intercept and resend attack by an eavesdropper Eve \cite{Xie2013}. Eve intercepts qudit 2 that 
Alice sends. She then prepares her own maximally entangled two-qudit state $\ket{c_e,r_e;s_e}_{3,4}$ from the set defined in 
~(\ref{eqn:crs}) and sends qudit 4 to Bob. Next she intercepts the returning qudit from Bob and measures qudits 3 and 4 in the basis $\{\ket{c'_e,r'_e;s_e}_{3,4}\}_{c'_e,r'_e=0}^{d-1}$. From her measurement outcome, she can 
infer $b$ except for probability $\frac{1}{d}$ when it is inconclusive. If her measurement  yields an inconclusive outcome, Eve forwards qudit 2 to Alice unmeasured. Otherwise Eve measures qudit 2 in the MUB labeled by $b$ and forward the measured qudit to Alice. Performing this attack, Eve learns the decoded signal $b$ and remains undetected.

\section{Secure communication}
In what follows we propose two secure communication protocols where  the message is encoded in the choice of the MUB measured by Bob. We prove the security of these communication protocols for various eavesdropping attacks. In these protocols, as well as in the original protocol, the actual numerical value of $c,r,s$ in the initially prepared state is irrelevant and we arbitrarily set $c=r=s=0$.

\subsection{First protocol}
For our first proposal, the communication can be made secure if Alice and Bob check that the qudits they share are indeed maximally entangled. To this end, Bob randomly choose a subset of his qudits to be used as `pre-test' qudits. He then choose a random index $b$ and measures each pre-test qudit in the MUB $\{\ket{m;b}\}_{m=0}^{d-1}$ and records the outcome $m$. He communicates the basis choice $b$ and outcome $m$ to Alice who then choose to measure her qudit in the MUB $\{\ket{m';a}\}_{m'=0}^{d-1}$ with a random index $a$. Since the MUBs are informationally complete \cite{ivanovic81}, Alice and Bob can determine whether or not that their qudits were indeed in the state $\ket{0,0;0}_{1,2}$ from the joint probability distributions of their measurement results. The non pre-test qudits are measured and decoded as in the original protocol and only the non-ambiguous outcomes from these qudits carry the communication signals. After Alice's measurements, Alice and Bob perform a `post-test' step. Alice randomly selects a fraction of her decoded result to check for maximum correlation with Bob's encoded signal. If Alice and Bob affirm from their pre-test results that their shared state are maximally entangled and from their post-test results that their correlations are maximum, we claim that they can be sure that their communication is secret.

In the following, we show that if Alice and Bob have the two-qudits maximally entangled state $\ket{0,0;0}_{1,2}$, then Eve cannot gain any information about the signal $b$. When Bob receives qudit 2, Alice and Bob have a pure maximally entangled state. This implies that Eve cannot share any quantum or classical correlation with Alice and Bob at this stage. After Bob measures qudit 2 and returns it to Alice, to Eve who is ignorant of the Bob's measurement outcome, qudit 2 is always in a completely mixed state regardless of the signal $b$. Hence this shows Eve cannot gain any information about $b$. However, Eve can still interact with qudit 2 to influence the outcome of Alice's measurement. But when Alice and Bob perform the post-test check on their decoded signals, any miscorrelations would alert them of Eve's interference. This proves our claim and concludes the first proposal. 

We should add that this protocol is a tomographic protocol \cite{bruss98,englert04}. Alice and Bob need to characterize the state they share before they can trust their channel. Obviously, any practical reconstruction scheme has unavoidable noise sources and the level of noise will have to be taken into account in a complete security proof. Such analysis, though is important for practical purposes, is out of the scope and aim of this paper. 

\subsection{Second protocol}
Our second proposal for secure communication is for Alice and Bob to utilize two different computational bases to define two families of MUBs. The choice of the second computational basis is immaterial as long as it is different from the original computational basis $\{\ket{n}\}_{n=0}^{d-1}$ and is not mutually unbiased to it. For example one can choose
\begin{equation}
\widehat{\ket{m}}= \sum_{n=0}^{d-1} \ket{n} h_{m,n}\;, m=0,1,\ldots,d-1\;,
\end{equation}
where the unitary matrix $h$ is the square root of the (complex) Hadamard matrix $H$ with matrix elements $H_{m,n}=e^{2 \pi i m n/d} / \sqrt{d}$. We define the hat MUBs as
\begin{equation}
\widehat {\ket{m;b}} = \frac{1}{\sqrt{d}} \sum_{n=0}^{d-1} \widehat{\ket{n}} \omega^{b n^2 -2 n m}\;,
\end{equation}
for $m,b=0,1,\ldots,d-1$, analogous to Eq.~(\ref{eqn:mub}).

In this proposal, Alice randomly creates either the state $\ket{0,0;0}_{1,2}$ or
\begin{equation}
\widehat{\ket{0,0;0}}_{1,2} =\frac{1}{\sqrt{d}} \sum_{n=0}^{d-1} \widehat{\ket{n}}_1\widehat{\ket{-n}}_2\;.
\end{equation}
Alice keeps qudit 1 and sends qudit 2 to Bob.  Bob randomly chooses to measure in one of the hat MUBs or in one of the original MUBs. His MUB choice, labeled by $b$, represents the signal to be sent. Bob returns the measured qudit to Alice. Alice then measures qudits 1 and 2 in her preparation basis, either $\{\ket{c,r;0}_{1,2}\}_{c,r=0}^{d-1}$ or $\{\widehat{\ket{c,r;0}}_{1,2}\}_{c,r=0}^{d-1}$ where
\begin{equation}
\widehat{\ket{c,r;0}}_{1,2} = \frac{1}{\sqrt{d}} \sum_{n=0}^{d-1}\widehat{ \ket{n}}_1\widehat{\ket{c-n}}_2 \omega^{ -2 r n}\;,
\end{equation}
with $c,r=0,1,\ldots, d-1$. After this stage, Bob announces his computational basis choice. For the cases where Alice and Bob used the same computational basis, Alice decodes the signal using~table (\ref{dec_tab}). The cases for which Alice and Bob's computational basis choices do not match or where the outcomes are inconclusive are discarded. They expect full correlations in the decoded signals. Alice and Bob select a random subset of the decoded signals to check that their correlations are indeed full.

We shall show that Eve cannot do an intercept and resend attack that renders the original protocol insecure. In the intercept and resend attack scenario we consider, Eve intercepts qudit 2 and creates one of two possible
maximally entangled state: either $ \ket{\psi}_{3,4}=\ket{0,0;0}_{3,4}$ or the hat version $\widehat{\ket{\psi}}_{3,4}=\widehat{\ket{0,0;0}}_{3,4}$. For our discussion, it is immaterial which of the state Eve actually prepares. Say that she prepares $ \ket{\psi}_{3,4}$. Eve keeps qudit 3 and forwards qudit 4 to Bob.

Bob performs his signal encoding by measuring qudit 4 in one of $2(d+1)$ MUBs. (There are $d+1$ MUBs defined in terms of the original computational basis and an additional $d+1$ MUBs in terms of the hat computational basis.) Thereafter Bob sends qudit 4 to Alice.

Eve intercepts qudit 4 on its way back. Suppose Bob measured in the $\{\ket{m;b} \}_{m=0}^{d-1}$ basis, the state Eve holds is then
\begin{equation}
\rho_{3,4}=\sum_{m=0}^{d-1} \ket{m;b}\braket{m;b}{\psi}\braket{\psi}{m;b}\bra{m;b}\;.
\end{equation}
This state comprises of an equal mixture of $d$ orthogonal pure states. Indeed $\rho_{3,4}$ can be written as
\begin{align}\label{rho34}
&\rho_{3,4}=\nonumber\\&\frac{1}{d}
\begin{cases}
 \sum_{r=0}^{d-1} \ket{0, r ; 0}\bra{0 ,  r ;0} &\mbox{for }b=\ddot{0}\\
 \sum_{c=0}^{d-1} \ket{c,-b c; 0}\bra{c, -b c;0} &\mbox{for }b=0,1,\ldots,d-1
\end{cases}\;,
\end{align}
which is diagonal in the maximally entangled basis $\{\ket{c,r;0}_{3,4}\}_{c,r=0}^{d-1}$.

Eve's eavesdropping strategy is now to measure qudits 3 and 4 in her preparation basis $\{\ket{c,r;0}_{3,4}\}_{c,r=0}^{d-1}$. (Had she prepared the state $\widehat{\ket{\psi}}_{3,4}$ instead, the strategy for her would then be to measure qudits 3 and 4 in the $\{\widehat{\ket{c,r;0}}_{3,4}\}_{c,r=0}^{d-1}$ basis.) Eve obtains full information about the decoded signal and remains undetected using the following strategy. If she gets the outcome $c=r=0$, her measurement is inconclusive and Eve returns qudit 2 to Alice unmeasured. Otherwise, Eve correctly infers Bob's encoding $b$ as $b=\ddot{0}$ if $c=0$ and $b=-r/c$ if $c\neq 0$. Eve now measures qudit 2 in the MUB labeled by $b$ in her chosen computational basis choice and forwards the measured qudit to Alice.

However, if Bob had measured in the hat basis $\{\widehat{\ket{m;b}} \}_{m=0}^{d-1}$ instead, the state Eve holds is now
\begin{equation}
\hat\rho_{3,4}=\sum_{m=0}^{d-1} \widehat{\ket{m;b}}\widehat{\bra{m;b}}{\psi}\rangle\langle{\psi}\widehat{\ket{m;b}}\widehat{\bra{m;b}}\;,
\end{equation}
which unlike $\rho_{3,4}$ is not diagonal in the $\{\ket{c,r;0 } \}_{c,r=0}^{d-1}$ maximally entangled basis. In this case when Eve measures the joint state $\hat{\rho}_{3,4}$ in her preparation basis $\{\ket{c,r;0 } \}_{c,r=0}^{d-1}$ and subsequently measures qubit 2 based on the outcome of that measurement, she would end up measuring qudit 2 in a different basis from Bob's. If after that, Alice measures in the same basis family as Bob's, due to Eve's forwarding a differently measured state, Alice will have a probability to obtain an outcome that was otherwise impossible without Eve's interference. This will reveal Eve's presence when Alice and Bob check their correlations.

We note that the security analysis here follows closely the reasoning of the security analysis of the original quantum cryptographic protocol by Bennett and Brassard \cite{bb84}. We concede that there may be other forms of more sophisticated attacks that Eve can perform without being detected. As the purpose of this work is to introduce this protocol and not to exhaust all possible attacks that Eve can perform, we leave that possibility and the discussion of a noisy channel for future work.

\section{Conclusion}
In conclusion, we have discussed a new possibility for secure encoding of information in quantum systems. Rather than encoding information in quantum states, we introduce two secure communication protocols where the information is encoded in measurement settings---in particular of complementary, incompatible measurements. In the first proposal, we show that any information leakage to Eve can be detected by Alice and Bob. In the second proposal, we show that Eve cannot do an intercept and resend attack and remain undetected.

\section{Acknowledgement}
SMA would like to acknowledge the support of the Australian Research Council Centre of Excellence for Quantum Computation and Communication Technology (project number CE110001027). AK would like to acknowledge the support of NSF Grants No. PHY-1212445.


\begin{thebibliography}{99}
\bibitem{bb84} 
C.H. Bennett and G. Brassard, {\it Proc. IEEE Int. Conf. on Computers Systems and Signal Processing pp. 175-179} (Bangalore IEEE, New York, 1984).
%Quantum cryptography: Public key distribution and coin tossing.
\bibitem{ekert91}
A.K. Ekert, Phys. Rev. Lett. {\bf 67}, 661 (1991).
%Quantum cryptography based on BellÕs theorem. 
\bibitem{bb92} 
C.H. Bennett and G. Brassard,  Phys. Rev. Lett. {\bf 68}, 3121 (1992).
%Quantum cryptography using any two non-orthogonal states.
\bibitem{bruss98}
D. Bru{\ss},  Phys. Rev. Lett. {\bf 81}, 3018 (1998).
%Optimal eavesdropping in quantum cryptography with six states.
\bibitem{englert04}
B.-G. Englert, D. Kaszlikowski, H.K. Ng, W.K. Chua, J. \v{R}eh\'{a}\v{c}ek and J. Anders,  quant-ph/0412075 (2004). 
%Efficient and robust quantum key distribution with minimal state tomography.
\bibitem{Kalev2013}  A. Kalev, A. Mann, and M. Revzen,  Phys. Rev Lett. {\bf 110}, 260502 (2013).
%Choice of measurement as the signal.
\bibitem{Xie2013}  D. Xie, A.M. Wang,  arXiv:1309.0312 (2013). 
%Secure communication with choice of measurement.
\bibitem{ivanovic81}  I. D. Ivanovic,  J. Phys. A {\bf 14}, 3241 (1981).
%Geometrical description of quantal state determination. 
\bibitem{tal}  S. Bandyopadhyay, P. O. Boykin, V. Roychowdhury and F. Vatan,  Algorithmica  {\bf 34}, 512 (2002).
%A new proof for the existence of mutually unbiased bases.
\bibitem{durt10} T. Durt, B.-G. Englert, I. Bengtsson and K. \.Zyczkowski,  Int. J. Quant. Info. {\bf 8}, 535 (2010).
%On mutually unbiased bases.
\bibitem{rev1} M. Revzen, Phys. Rev. A {\bf 81}, 012113 (2010).
%Maximally entangled states via mutual unbiased collective bases.
\bibitem{rev2} M. Revzen,  J. Phys. A {\bf 46}, 075303 (2013).
%Maximal entanglement, collective coordinates and tracking the King.
\end{thebibliography}
\end{document}